\documentclass[twocolumn]{aastex63}
\usepackage[T1]{fontenc}

\usepackage[utf8]{inputenc}
\usepackage{enumitem}
\usepackage{float} 
\usepackage{gensymb}
\usepackage{amsmath}
\usepackage{textcomp}
\usepackage{verbatim}
\usepackage{tabularx}
\usepackage{placeins}
\usepackage{multirow}

\newcommand{\kms}          {\mbox{${\rm km~s^{-1}}$}}

\newcommand{\cc}           {\mbox{${\rm cm^{-3}}$}}
\newcommand{\e}            {\mbox{$^{-1}$}}
\newcommand{\ee}           {\mbox{$^{-2}$}}

\def\cm2{\mbox{${\rm cm^{-2}}$}}
\def\h2{\mbox{${\rm H}_2$}}
\def\nh2{\mbox{$n_{\rm H_2}$}}
\def\Nh2{\mbox{$N_{{\rm H}_2}$}}
\def\Mh2{\mbox{$M_{{\rm H}_2}$}}

\def\farcs{\hbox{$.\!\!^{''}$}}

\def\simgt{\lower.5ex\hbox{$\; \buildrel > \over \sim \;$}}
\def\simlt{\lower.5ex\hbox{$\; \buildrel < \over \sim \;$}}

\def\13CO{$^{13}$CO}
\def\C18O{C$^{18}$O}
\def\H2{H$_2$}

\def\startfigcap{\vspace*{2.0\baselineskip}\bgroup\leftskip 0.45in\rightskip 0.45in\small}
\def\endfigcap{\par\egroup\vspace*{2.0\baselineskip}}

\def\startfigcapside{\vspace*{2.0\baselineskip}\bgroup\leftskip 4.0in\rightskip 0.1in\small}
\def\endfigcapside{\par\egroup\vspace*{2.0\baselineskip}}

\usepackage{color}
\definecolor{CornflowerBlue}{rgb}{0.39,0.58,0.93}
\definecolor{TitleBrown}{rgb}{0.60,0.55,0.52}   
\definecolor{SectionRed}{rgb}{0.55,0.15, 0.17}
\definecolor{royalblue}{rgb}{0.25,0.41,0.88}
\definecolor{pureblue}{rgb}{0.0,0.0,1.0}
\definecolor{darkblue}{rgb}{0.05, 0.05, 0.45}
\definecolor{firebrick1}{rgb}{1.00,0.19,0.19}

\makeatletter
\newcommand{\smallsym}[2]{#1{\mathpalette\make@small@sym{#2}}}
\newcommand{\make@small@sym}[2]{%
  \vcenter{\hbox{$\m@th\downgrade@style#1#2$}}%
}
\newcommand{\downgrade@style}[1]{%
  \ifx#1\displaystyle\scriptstyle\else
    \ifx#1\textstyle\scriptstyle\else
      \scriptscriptstyle
  \fi\fi
}
\makeatother

\usepackage[version=4]{mhchem}
\shorttitle{Disks, Infall, and a Fossil Outburst in Oph\,IRS43}
\shortauthors{Narayanan, Williams, \& eDisk Team}

\begin{document}
\title{Early Planet Formation in Embedded Disks (eDisk) X:\\
Compact Disks, Extended Infall, and a Fossil Outburst in the Class I Oph\,IRS43 Binary}

\correspondingauthor{Jonathan Williams}
\email{jw@hawaii.edu}

\author[0000-0002-0244-6650]{Suchitra Narayanan}
\altaffiliation{National Science Foundation Graduate Research Fellow}
\affiliation{Institute for Astronomy, University of Hawai`i at Mānoa, 2680 Woodlawn Dr., Honolulu, HI 96822, USA}

\author[0000-0001-5058-695X]{Jonathan P. Williams}
\affiliation{Institute for Astronomy, University of Hawai`i at Mānoa, 2680 Woodlawn Dr., Honolulu, HI 96822, USA}

\author[0000-0002-6195-0152]{John J. Tobin}
\affiliation{National Radio Astronomy Observatory, 520 Edgemont Rd., Charlottesville, VA 22903, USA}

\author[0000-0001-9133-8047]{Jes K. J{\o}rgensen}
\affiliation{Niels Bohr Institute, University of Copenhagen, {\O}ster Voldgade 5-7, 1350, Copenhagen K, Denmark}

\author[0000-0003-0998-5064]{Nagayoshi Ohashi}
\affiliation{Academia Sinica Institute of Astronomy \& Astrophysics, 11F of Astronomy-Mathematics Building, AS/NTU, No.1, Sec. 4, Roosevelt Rd, Taipei 10617, Taiwan, R.O.C.}

\author[0000-0001-7233-4171]{Zhe-Yu Daniel Lin}
\affiliation{University of Virginia, 530 McCormick Rd., Charlottesville, Virginia 22904, USA}

\author[0000-0002-2555-9869]{Merel L.R. van 't Hoff}
\affil{Department of Astronomy, University of Michigan, 1085 S. University Ave., Ann Arbor, MI 48109-1107, USA}

\author[0000-0002-7402-6487]{Zhi-Yun Li}
\affiliation{University of Virginia, 530 McCormick Rd., Charlottesville, Virginia 22904, USA}

\author[0000-0002-9912-5705]{Adele L. Plunkett}
\affiliation{National Radio Astronomy Observatory, 520 Edgemont Rd., Charlottesville, VA 22903, USA}

\author[0000-0002-4540-6587]{Leslie W. Looney}
\affiliation{Department of Astronomy, University of Illinois, 1002 West Green St, Urbana, IL 61801, USA}

\author[0000-0003-0845-128X]{Shigehisa Takakuwa}
\affiliation{Department of Physics and Astronomy, Graduate School of Science and Engineering, Kagoshima University, 1-21-35 Korimoto, Kagoshima,Kagoshima 890-0065, Japan}
\affiliation{Academia Sinica Institute of Astronomy \& Astrophysics, 11F of Astronomy-Mathematics Building, AS/NTU, No.1, Sec. 4, Roosevelt Rd, Taipei 10617, Taiwan, R.O.C.}

\author[0000-0003-1412-893X]{Hsi-Wei Yen}
\affiliation{Academia Sinica Institute of Astronomy \& Astrophysics, 11F of Astronomy-Mathematics Building, AS/NTU, No.1, Sec. 4, Roosevelt Rd, Taipei 10617, Taiwan, R.O.C.}

\author[0000-0002-8238-7709]{Yusuke Aso}
\affiliation{Korea Astronomy and Space Science Institute, 776 Daedeok-daero, Yuseong-gu, Daejeon 34055, Republic of Korea}

\author[0000-0002-8591-472X]{Christian Flores}
\affiliation{Academia Sinica Institute of Astronomy \& Astrophysics, 11F of Astronomy-Mathematics Building, AS/NTU, No.1, Sec. 4, Roosevelt Rd, Taipei 10617, Taiwan, R.O.C.}

\author[0000-0003-3119-2087]{Jeong-Eun Lee}
\affiliation{Department of Physics and Astronomy, Seoul National University, 1 Gwanak-ro, Gwanak-gu, Seoul 08826, Korea }

\author[0000-0001-5522-486X]{Shih-Ping Lai}
\affiliation{Institute of Astronomy, National Tsing Hua University, No. 101, Section 2, Kuang-Fu Road, Hsinchu 30013, Taiwan}
\affiliation{Center for Informatics and Computation in Astronomy, National Tsing Hua University, No. 101, Section 2, Kuang-Fu Road, Hsinchu 30013, Taiwan}
\affiliation{Department of Physics, National Tsing Hua University, No. 101, Section 2, Kuang-Fu Road, Hsinchu 30013, Taiwan}
\affiliation{Academia Sinica Institute of Astronomy \& Astrophysics, 11F of Astronomy-Mathematics Building, AS/NTU, No.1, Sec. 4, Roosevelt Rd, Taipei 10617, Taiwan, R.O.C.}

\author[0000-0003-4022-4132]{Woojin Kwon}
\affiliation{Department of Earth Science Education, Seoul National University, 1 Gwanak-ro, Gwanak-gu, Seoul 08826, Republic of Korea}
\affiliation{SNU Astronomy Research Center, Seoul National University, 1 Gwanak-ro, Gwanak-gu, Seoul 08826, Republic of Korea}

\author[0000-0003-4518-407X]{Itziar de Gregorio-Monsalvo}
\affil{European Southern Observatory, Alonso de Cordova 3107, Casilla 19, Vitacura, Santiago, Chile}

\author[0000-0002-0549-544X]{Rajeeb Sharma}, 
\affiliation{Niels Bohr Institute, University of Copenhagen, \O ster Voldgade 5-7, 1350, Copenhagen K, Denmark}

\author[0000-0002-3179-6334]{Chang Won Lee}
\affiliation{Korea Astronomy and Space Science Institute, 776 Daedeok-daero, Yuseong-gu, Daejeon 34055, Republic of Korea}
\affiliation{Division of Astronomy and Space Science, University of Science and Technology, 217 Gajeong-ro, Yuseong-gu, Daejeon 34113, Republic of Korea}

\begin{abstract}
We present the first results from the Early Planet Formation in Embedded Disks (eDisk) ALMA Large Program toward Oph\,IRS43, a binary system of solar mass protostars. The 1.3 mm dust continuum observations resolve a compact disk, $\sim 6$\,au radius, around the northern component and show that the disk around the southern component is even smaller, $\lesssim 3$\,au. CO, \13CO, and \C18O\ maps reveal a large cavity in a low mass envelope that shows kinematic signatures of rotation and infall extending out to $\sim 2000$\,au. An expanding CO bubble centered on the extrapolated location of the source $\sim 130$ years ago suggests a recent outburst. Despite the small size of the disks, the overall picture is of a remarkably large and dynamically active region. 
\end{abstract}



\section{Introduction}
More than half of pre-main sequence stars are binaries \citep{Raghavan_binary_survey}. Such systems are thought to form through the fragmentation of dense cores and/or massive disks, and the interaction between the young stars may significantly affect their surroundings and evolution \citep{Offner2010}. However, disks can persist around each individual star and/or the system and ultimately turn into stable planetary systems \citep{2015pes..book..309T}. The study of young binary systems is therefore important both for a more complete picture of star and planet formation in general and also to extend the range of physical conditions for testing models.

The focus of this paper is on the protostellar binary system, Oph\,IRS43 (also known as YLW\,15 and GY\,265), that was observed as part of the Atacama Large Millimeter Array (ALMA) Large Program, Early Planet Formation in Embedded Disks (hereafter eDisk). This program, described in \cite{Ohashi_edisk_overview}, observed 12 Class 0 and 7 Class I protostars at high spatial resolution (0\farcs04) with the primary goal of studying the properties of their accompanying disks.

Oph\,IRS43, hereafter IRS43, is a Class I embedded protostellar binary located in the L1688 region of the Ophiuchus molecular cloud complex. We adopt a distance to the source of 137.3\,pc based on Very Long Baseline Array (VLBA) parallax measurements of 12 other (single) young stellar objects in L1688 \citep{OrtizLeon2017}.
The combined bolometric luminosity and temperature of the system are $L_{\rm bol}=4.15\,L_\odot, ~T_{\rm bol}=193$\,K \citep[see][]{Ohashi_edisk_overview}.

\citet{Girart2000} first showed that IRS43 was a $0\farcs6$ binary from centimeter wavelength observations with the Very Large Array (VLA) carried out in 1989. We follow their naming convention and designate the northern source, VLA1, and the southern source, VLA2.
The latter is much brighter in the near-infrared \citep{2007A&A...476..229D} and spectroscopy shows that it is a heavily extincted, $A_{\rm V}=40$\,mag, cool star with a KIV/V spectral type corresponding to an effective temperature $\sim 4300$\,K. It has a strong continuum excess (veiling = 3.0 at $2.2\,\mu$m) that indicates a high accretion rate, $\sim 10^{-6}\,M_\odot$\,yr\e, and is rapidly rotating at a rate, $v\sin i\sim 50$\,\kms, that is typical of embedded protostars \citep{Greene2002}.

Subsequent VLA imaging over 12 years revealed a common proper motion of 24 milli-arcsecond yr\e and relative orbital motion of the binary \citep{curiel}. ALMA observations of the source were presented by \citet{Brinch2016} who used the extended time baseline and archival VLA data to extend the astrometry and determine an orbital solution. They found that the motions are consistent with a circular orbit in or near the plane of the sky with a period of 450 years and a total mass of $2.01\pm0.47\,M_\odot$ with an equal mass ratio, i.e. two solar mass stars. These astrometric constraints on the proper motion and total mass are essential to our interpretation of the eDisk data in \S\ref{sec:results}.

The first millimeter interferometric observations of IRS43 were made using the Submillimeter Array (SMA) by \cite{2009jes} and \cite{Brinch2013}, revealing that IRS43 has strong lines on top of a weak continuum with an extended, flattened structure in HCO$^+$ but the $1.7''\times 1.4''$ resolution was too low to resolve the binary. The \cite{Brinch2016} ALMA observations were the highest resolution measurements of this system to date, $0\farcs2$, and clearly separated the disks around each protostar but they remained unresolved with an upper limit to their radius of $\sim 20$\,au. However, the HCO$^+$ and HCN line data suggested that the larger scale flattened structure was rotating around the binary and that the misalignment between the stellar orbits and circumbinary material testified to a turbulent origin.

IRS43 is also a well-known X-ray source that undergoes energetic flares every $\sim 20$\,hr quasi-periodically due most likely to a strong star-disk interface \citep{Montmerle2000}. It stands out for having undergone the brightest ``superflare'' ever witnessed in T Tauri stars, when in 1995 its X-ray luminosity peaked at $L_{\rm X}\sim 10-100\,L_\odot$ and outshone the entire system at all other wavelengths for a couple of hours \citep{Grosso1997}. Although the positional accuracy of the data was unable to determine whether the source was VLA1 or VLA2, such a superflare must have been powered by a massive accretion event and would have been accompanied by full ionization of the surroundings within a few tenths of an au. Our eDisk observations suggest a different, though perhaps related, outburst event occurred $\sim 130$\,yr ago.

The rest of the paper is organized as follows: \S\ref{sec:obs} describes the ALMA observations, data reduction and imaging procedure. \S\ref{sec:results} presents the results separated into subsections focused on the 1.3\,mm dust continuum, the molecular line data showing the kinematics of the system, and an expanding bubble that signposts a significant recent outburst. \S\ref{sec:discussion} discusses the implications of our findings and \S\ref{sec:summary} summarizes the paper.

\section{Observations and Data Reduction}\label{sec:obsanddataredux}
\subsection{Observations}\label{sec:obs}
The ALMA observations used in this work were taken in Cycle 7 as part of program 2019.1.00261.L (PI: N. Ohashi). Detailed information on the configurations, spectral setup and targeted lines, and the other observed targets are in \cite{Ohashi_edisk_overview}. In brief, the IRS43 observations were carried out in 5 execution blocks (EB) between May and October 2021 with between 41 and 46 antennas in a compact and extended configuration (C43-5 and C43-8, respectively) with baselines extending over 15-11500\,m. The total on-source integration time was 170\,minutes in a single Band 6 spectral setting. Here we present results from the 234\,GHz continuum and the CO, \13CO, \C18O, SO, and H$_2$CO lines at 219-230\,GHz.

\subsection{Calibration}\label{sec:reduction}
The data were calibrated using the Common Astronomy Software Applications ({\tt CASA}) package \citep{McMullin2007} version~{6.2.1}.
To ensure uniformity in the eDisk data products and comparison of structures and kinematics across the sample, the project team created a specific eDisk data reduction routine\footnote{All of the data reduction (i.e., self-calibration and imaging) can be found at \url{http://github.com/jjtobin/edisk}.}, that builds on the calibration strategy developed for the Disk Substructures at High Angular Resolution Project (DSHARP) ALMA Large Program \citep{DSHARP}.
That general reduction procedure is described in \cite{Ohashi_edisk_overview} but an additional step (the ``two-pass'' method) was required for 5 of the 19 sources, including IRS43, where some of the delivered data had high phase decorrelation which we describe below.

\subsubsection{Combining multiple datasets with high phase noise}\label{sec:cal}
\begin{figure}
\centering
\includegraphics[width=\columnwidth]{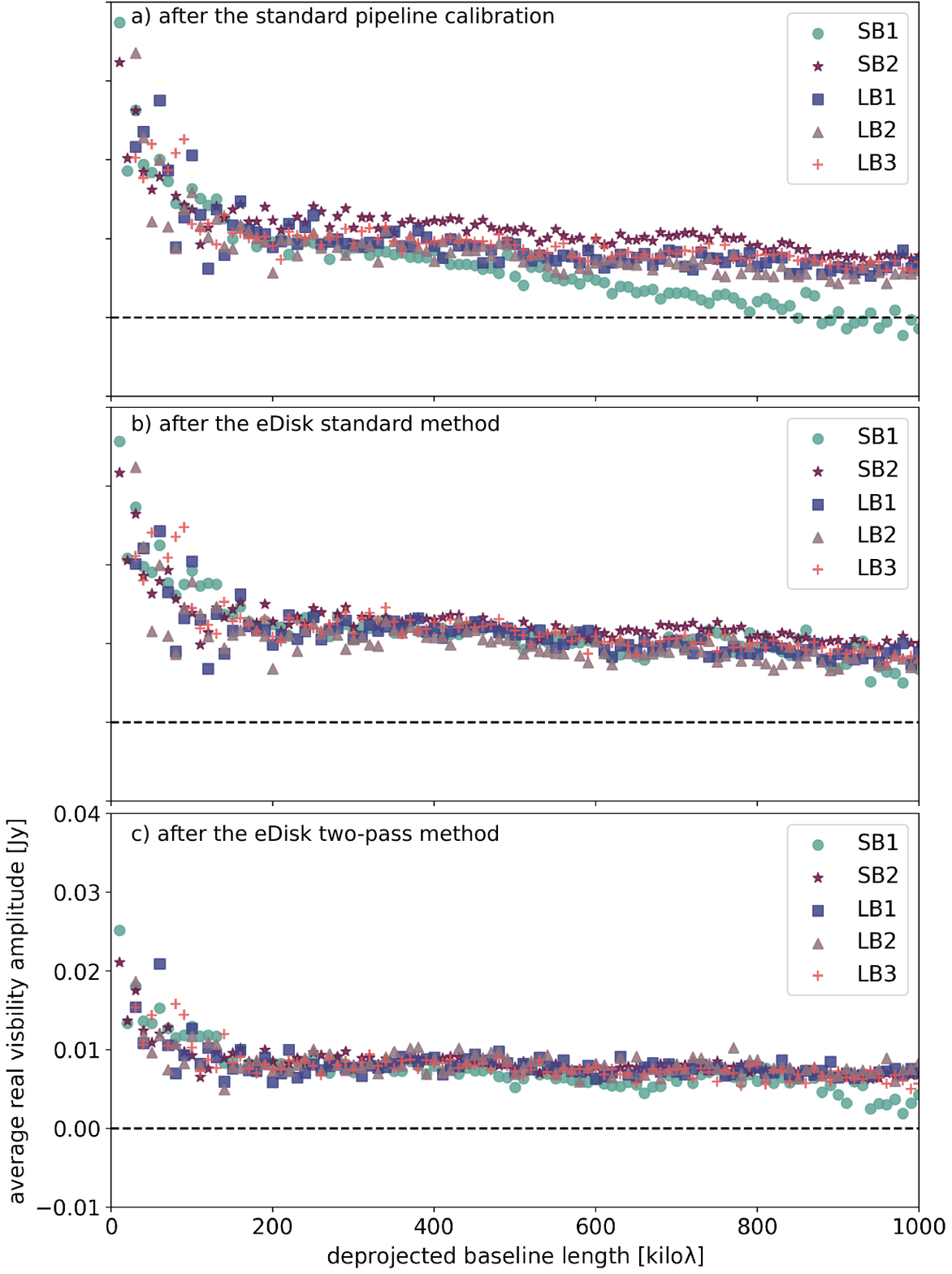}
\caption{\textit{Top:} Visibility amplitudes as a function of \textit{uv}-distance after the standard pipeline calibration. Note that short-baseline 1 (SB1) decreases anomalously due to loss of coherence on long baselines. 
\textit{Middle:} Same as top panel but after the standard self-calibration procedure  \cite[for detailed description see][]{Ohashi_edisk_overview}.
\textit{Bottom:} Same as top panel but after joint self-calibration using the two-pass method. The procedure is described in Section \ref{sec:cal}. For these data, the long-baseline 3 (LB3) was chosen as the reference, or best, dataset to which all the other datasets were scaled.}
\label{fig:scaling}
\end{figure}

Each execution block was first passed through the standard ALMA data reduction pipeline (version 2021.2.0.128) to remove atmospheric and instrumental effects using the quasar calibrators followed by self-calibration on the source itself to increase the signal-to-noise ratio. The top panel of Figure~\ref{fig:scaling} shows the real part of the source visibility amplitudes for each dataset across a range of overlapping baseline lengths. There is some scatter due to time variability in the calibrators and/or source. Notably, there is a significant loss of flux at long baselines in the first short-baseline track (SB1) due to decorrelation caused by high phase noise.
The middle panel of Figure~\ref{fig:scaling} shows what the visibility amplitudes would be had we followed the standard eDisk reduction procedure where the fluxes are normalized to a common scale first and then self-calibrated (see \cite{Ohashi_edisk_overview}).
In the ``two-pass'' method, however, we first run the self-calibration to completion without scaling the fluxes.
We then determined which EB had the the highest quality data based on the smoothness of the visibility-amplitude profiles and quality of the images made on a per-EB basis, and selected this to be the reference EB. The reference EB showed the least amount of decorrelation and lowest level of imaging artifacts due to phase noise. For IRS43, this was the third long-baseline (LB3). We subsequently rescaled the other EBs to the same average amplitude for $uv-$distances out to $800\,{\rm k}\lambda$, and re-ran the data reduction script from the start with the pre-determined scaling factors. The result of this second pass through the calibration process, shown in the bottom panel of Figure~\ref{fig:scaling}, has a much more uniform flux profile and reduced scatter across all datasets and baselines (consistent with a $\sim 10$\% absolute flux calibraiton uncertainty) which enables a higher quality image to be produced from their combination.

\subsection{Imaging}
Due to the range of spatial scales and surface brightness of the various features, we imaged the data using different baseline ranges and visibility weightings. To provide an overall view and show the faint, extended continuum emission, we used the short-baseline data only (hereafter SB) and natural weighting (Briggs ${\tt robust}=2$) with a correspondingly relatively large beam size $0\farcs31\times0\farcs24$ and low continuum RMS noise levels, 0.015\,mJy\,beam\e.
For the highest resolution view of the disks, we combined the short- and long-baseline data (hereafter SB+LB) and weighted more toward the longer baselines (Briggs weighting ${\tt robust}=-1$ or $-0.5$) to achieve much smaller beam sizes \mbox{$\sim 0\farcs04$} but at the expense of higher continuum RMS  \mbox{$\sim 0.05$}\,mJy\,beam\e.
The line maps were all produced from the SB+LB data using a $2000\,{\rm k}\lambda$ taper and intermediate weighting (Briggs ${\tt robust}=0.5$) with typical beam sizes $\sim 0\farcs15$ and RMS levels $\sim 1.5$\,mJy\,beam\e\,\kms.
For the weaker lines, we smoothed the data to a beam size of $0\farcs25$.

\section{Results}\label{sec:results}
\subsection{Continuum data}\label{sec:continuum}

\begin{figure*}[ht]
\centering
\includegraphics[width=\textwidth ]{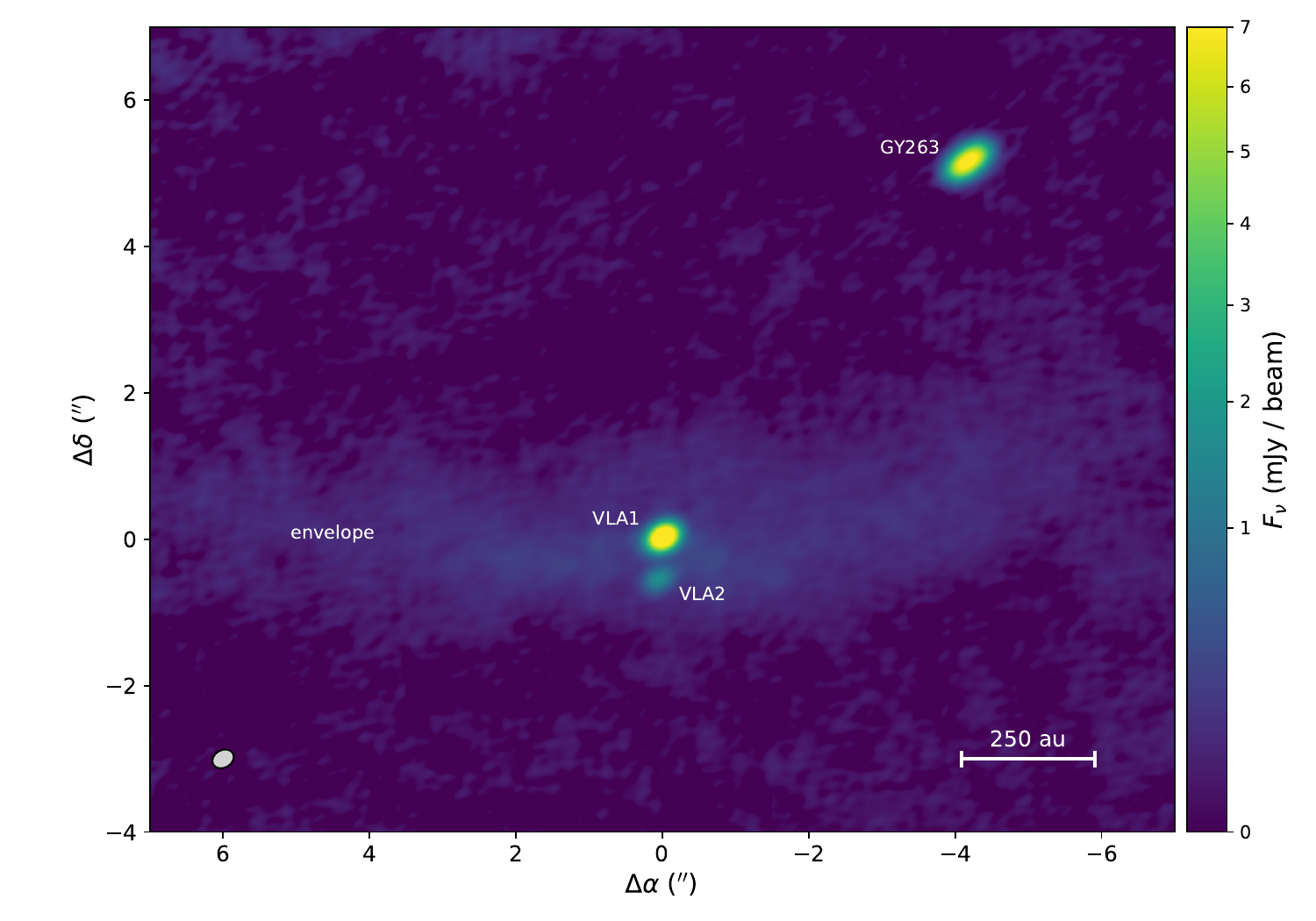}
\caption {1.3\,mm continuum image using only the short baseline data with a log stretch to emphasize the faint, extended emission around the IRS43 binary. The coordinates are arcsecond offsets from the position of VLA1 and the synthesized beam on the bottom left corresponds to 0\farcs31 $\times$ 0\farcs24.}
\label{fig:continuum_overall}
\end{figure*}

\subsubsection{Overall view}\label{sec:extended_continuum}
A map of the large scale 1.3\,mm continuum emission toward the region, optimized to highlight low surface brightness features more than high resolution, is shown in Figure\,\ref{fig:continuum_overall}.
There are 3 compact sources corresponding to disks around each member of the IRS43 binary and also the infrared source, GY\,263. We present higher resolution images of these in the following subsection. There is also an elongated, slightly curved structure that extends $\sim 7'' \approx 10^3$\,au on either side of the binary.
The line data, presented in \S\ref{sec:line}, reveal this to be a rotating, infalling envelope.
The flux density of this extended continuum emission (without the disks) is 36\,mJy, which corresponds to a total dust mass of only $10\,M_\oplus$ at a temperature of 30\,K derived from H$_2$CO line ratios. Assuming an ISM gas-to-dust ratio of 100, the implied total mass is $3\times 10^{-3}\,M_\odot$.
Note, however, that the envelope size is greater than the maximum recoverable size of the observations, $\sim 3''$, so this envelope mass measurement should be considered a lower limit.

\subsubsection{Circumstellar disks}\label{sec:circumstellar}
\begin{figure}
    \centering
    \includegraphics[width=\columnwidth]{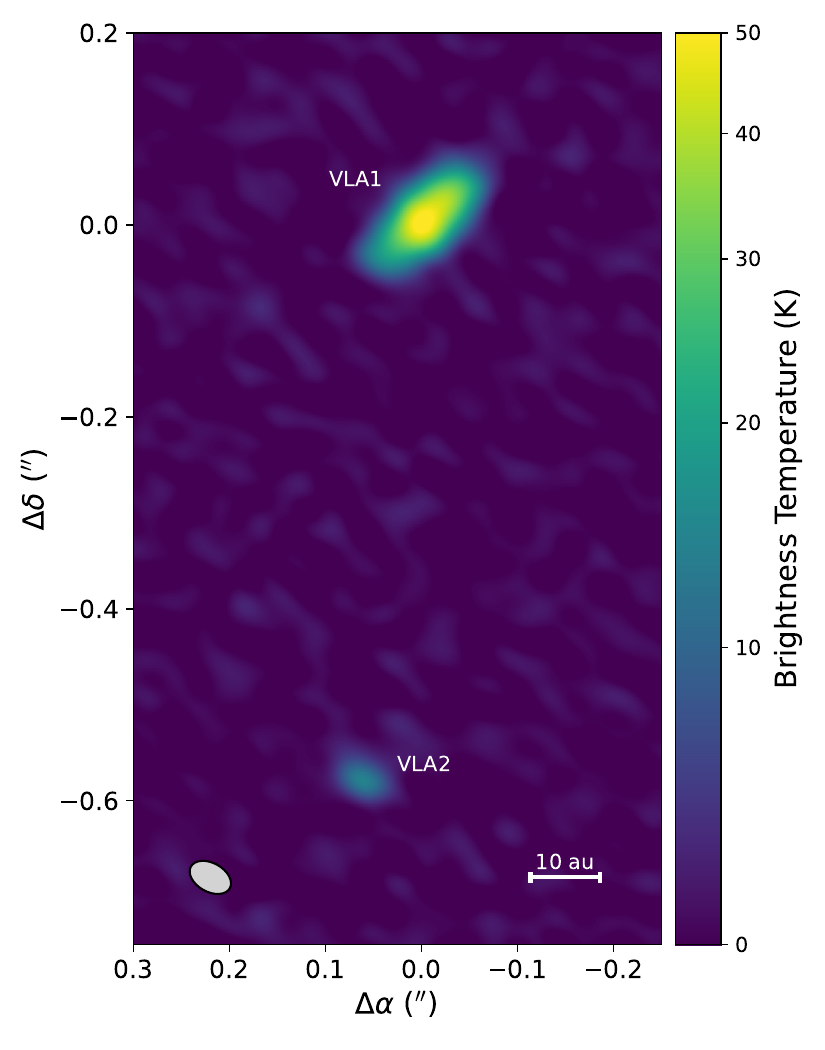}
    \caption{High resolution 1.3\,mm continuum image of the compact dusty disks around each member of the IRS43 binary. This was produced by strongly weighting the long baseline visibilities and is shown on an Asinh intensity scale. The beam size is $0\farcs046\times 0\farcs030$ ($\sim 6\,{\rm au}\times4$\,au).}
    \label{fig:continuum_binary}
\end{figure}

Including the long-baseline data and using different weighting mechanisms, we can zoom into each of the three disks. Figure \ref{fig:continuum_binary} shows the two disks in the IRS43 binary from the SB+LB data weighted heavily to long baselines (Briggs ${\tt robust}=-1$).
The disks are extremely compact. Their positions and sizes were determined from Gaussian fits in the image plane using the {\tt CASA imfit} routine and listed in Table \ref{tab:imfitresults}.
VLA1 is resolved in both major and minor axes, with a deconvolved FWHM size of 12\,au,
and, assuming a flat disk, has an inclination inferred from the arc-cosine of their ratio equal to $78\arcdeg$.
VLA2 is just barely resolved along the beam minor axis with a remarkably small deconvolved FWHM size of 2.3\,au. The source is faint so we checked different weighting schemes and found that it was detected at slightly higher signal-to-noise but also resolved in a slightly larger beam with a similar deconvolved size (2.7\,au) with Briggs ${\tt robust}=-0.5$.
To our knowledge, this is the smallest resolved protostellar disk in the ALMA literature to date.

The flux densities are 10.6 mJy and 1.1 mJy for VLA1 and VLA2,  respectively. Using the simplest assumptions of optically thin, isothermal emission, these convert to dust masses $3.6\,M_\oplus$ and $0.38\,M_\oplus$ at a temperature $T_{\rm dust}=30$\,K appropriate for Class I sources \citep{Tobin2015} and a dust opacity $\kappa_\nu = 2.3\,{\rm cm^2\,g^{-1}}$ \citep{beckwithB}. These are likely lower limits, especially for VLA1, since the high brightness temperature indicates that the emission is in fact optically thick.

\begin{deluxetable*}{cccccccc}
\tablenum{3}
\tablecaption{Gaussian fits to the continuum sources\label{tab:imfitresults}}
\tablewidth{0pt}
\tablehead{
\colhead{Source} & \colhead{ICRS R.A. } & \colhead{ICRS Dec. } & \colhead{Deconvolved FWHM, PA} &
\colhead{Peak $I_{\nu}, T_{\rm b}$} & \colhead{$F_{\nu}$} & \colhead{incl} & Dust Mass \\[-2mm]
\nocolhead{dummy} & \colhead{[h m s]} & \colhead{[d m s]} &
\colhead{[mas,\arcdeg]} & \colhead{[mJy beam$^{-1}$, K]} & \colhead{[mJy]} & \colhead{[\arcdeg]} & \colhead{[$M_\oplus$]}
}
\startdata
VLA1 & 16:27:26.906 & $-$24:40:50.81 & 89$\times$18, 133.5 $\pm$ 1.1 & 3.30, 62 & 10.62 & 78 & 3.4  \\
VLA2 & 16:27:26.911 & $-$24:40:51.40 & 19$\times$16,  62.5 $\pm$ 6.3 & 0.88, 20 &  1.08 & 32 & 0.34 \\
\enddata
\end{deluxetable*}

\begin{figure}
    \centering
     \includegraphics[width=\columnwidth]{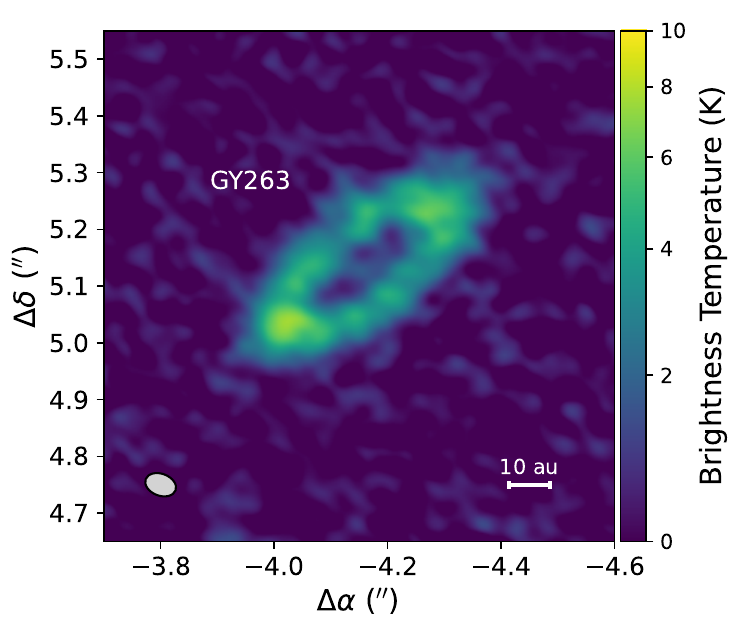}
    \caption{Zooming in on the continuum image toward the infrared source GY\,263 with a beam size of $0\farcs056\times 0\farcs039$ and an Asinh intensity scale. There is a well resolved central cavity showing that this is a small transition disk.}
    \label{fig:GY263}
\end{figure}

The third disk lies around the known Class II source, GY 263 \citep{Allen2002},
6\farcs6 (900\,au) to the North-West of IRS43. Figure~\ref{fig:GY263} shows a high resolution image from the SB+LB data with Briggs ${\tt robust}=-0.5$.
Although this disk was not the target of the eDisk program, it is nonetheless interesting in its own right as we serendipitously find a central hole.
The disk flux of 13.1\,mJy implies a dust mass, $4.5\,M_\oplus$, and the ring has a radius of $0\farcs17 = 23$\,au with an inclination of $24^\circ$ at a position angle of $130^\circ$.
This tells us that GY 263 is a small transition disk \citep[e.g.,][]{vanderMarel2022} while also demonstrating both the ability of the eDisk data to image small scale disk features and the clear visual differences with the two embedded disks.
Nevertheless, given the large projected distance between GY 263 and IRS43, and because we do not see any signs of interaction in the line data or in archival datasets, we consider the sources to be essentially physically independent and do not discuss GY 263 further here.

\subsection{Line data}\label{sec:line}
Spectral line emission is detected predominantly from $v_{\rm LSR}=-7$ to 2\,\kms\ and 6 to 15\,\kms\ in CO and its isotopologues, H$_2$CO, and SO.
There is very little signal around the central velocities of the system, 2 to 6\,\kms, in CO and \13CO. The most likely cause is spatial filtering of large scale emission, although there may also be some absorption by an intervening cold molecular layer.
The actual central velocity of each source cannot be precisely determined as we do not detect any line emission that can be clearly associated with either of the two disks.
However, from the more red- and blue-shifted parts of the spectra, we can study the kinematics of the envelope seen in the extended continuum.

\subsubsection{Overall view}\label{sec:envelope}

\begin{figure*}[ht]
\centering
\includegraphics[width=\textwidth]{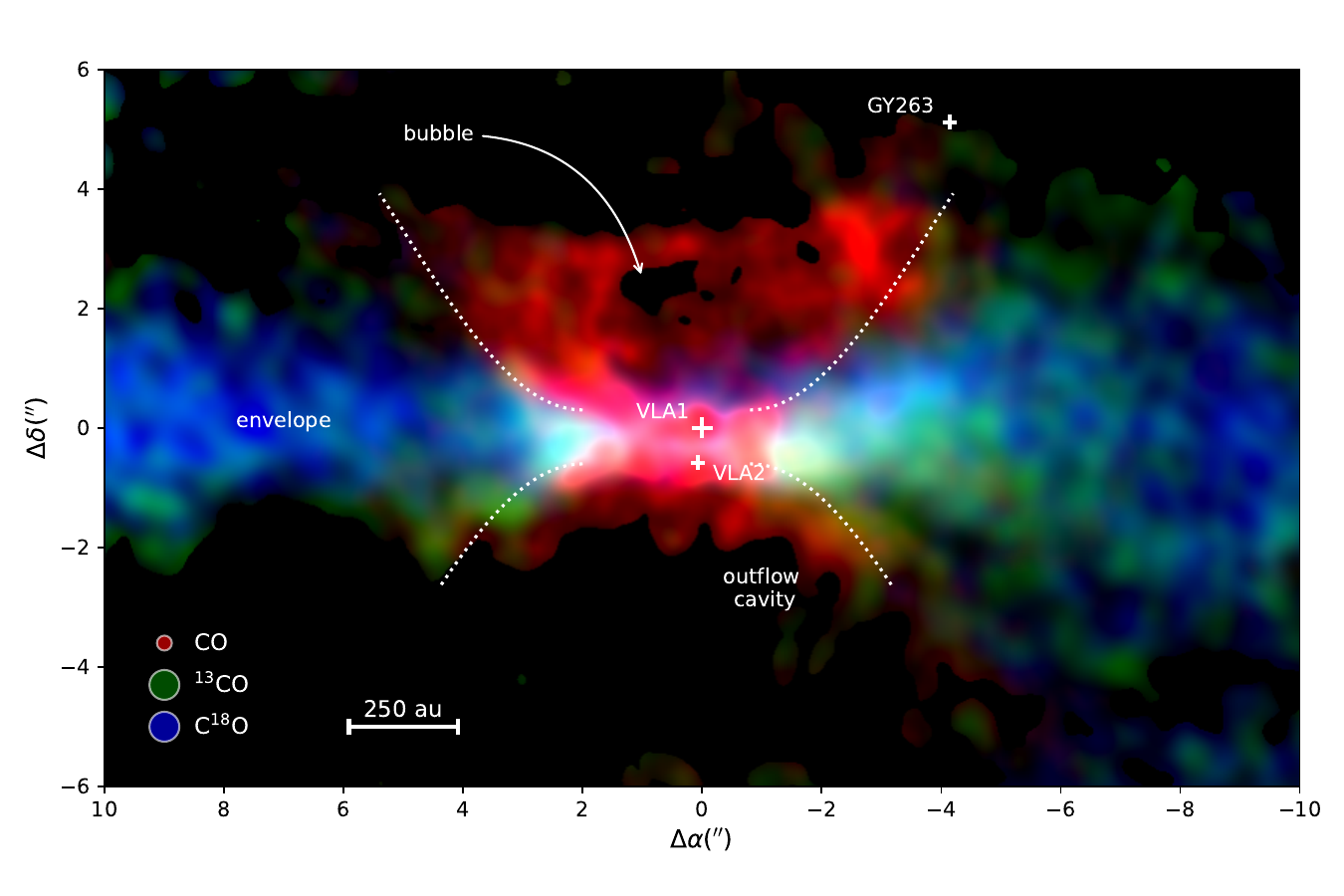}
\caption{Velocity integrated maps of the CO, \13CO, and \C18O\ emission (from -3 to +11 km/s) in red, green, and blue, respectively. The maps are centered on VLA1 and each autoscaled from their minimum to maximum and shown on a linear scaling. The CO map has been smoothed to 0\farcs25 and the isotopologues to 0\farcs5 to enhance the extended structures. The positions of the three disks are shown by the white crosses and other prominent features are labeled.}
\label{fig:envelope}
\end{figure*}

The IRS43 protostellar binary lies at the center of a molecular filament or envelope that we detect in CO 2--1 and its isotopologues \13CO\ and \C18O, as well as H$_2$CO $3_{0,3}-2_{0,2}$ and SO $6_5-5_4$. A 3-color CO, \13CO, and \C18O\ moment 0 map is shown in Figure~\ref{fig:envelope}. The overlay of the 3 lines gives a better representation of the molecular gas structure than any single line due to the range of column densities that they trace and the high level of absorption and interferometric filtering of large scale structure.

The faint extended dust emission in Figure~\ref{fig:continuum_overall} is best seen in the CO isotopologues because the emission from the optically thick CO line is relatively uniform resulting in a weak interferometric response. The total integrated \13CO\ emission over the entirety (760 square arcseconds) of the mapped structure is 39 Jy\,\kms\ which corresponds to a mean column density of $1.1\times 10^{15}$\,cm\ee\ at 30\,K and a total gas mass $\sim 4\times 10^{-4}\,M_\odot$ for an abundance $[^{13}{\rm CO}]/[{\rm H}_2]=2\times 10^{-6}$ \citep{2013MNRAS.431.1296R},
which in good agreement with the dust-derived mass.
The CO emission is more prominent along the bright rims of a cavity that extends north (most prominently) and south of the circumbinary envelope. There is a hole in the northern structure that is more apparent in the channel maps which we discuss further in \S\ref{sec:bubble}.

There is a slight enhancement of CO toward VLA1 and VLA2 in the moment map but, partially due to the strong obscuration or filtering around the central velocities, we were unable to identify line emission in any molecule or transition that could be clearly identified withither disk.
\citet{2014ApJ...788...59W} calculated the \13CO\ and \C18O\ emission for a grid of disk models with varying masses and abundances, marginalizing over size, surface density, and temperature profiles.
From the non-detections here, we estimate a $3\sigma$ upper limit to the gas mass of $0.3\,M_{\rm Jup}$ for each disk. This is about 30 and 300 times the dust mass for VLA1 and VLA2, respectively.

The observing passband contains two additional H$_2$CO lines, $3_{2,1}-2_{2,0}$ and $3_{2,2}-2_{2,1}$, that were weakly detected in the envelope. The relative strength of these higher excitation transitions constrain the gas temperature. 
The spatially and velocity integrated ratio of line strengths for the ${3_{0,3}-2_{0,2}}/{3_{2,1}-2_{2,0}}$ transitions is $7\pm1.5$, from which we derive an average rotational temperature of $28\pm 4$\,K assuming LTE and using the radiative transfer code RADEX \citep{vanderTak2007}.

\subsubsection{Kinematics}\label{sec:kinematics}

\begin{figure*}[ht]
\includegraphics[width=\textwidth]{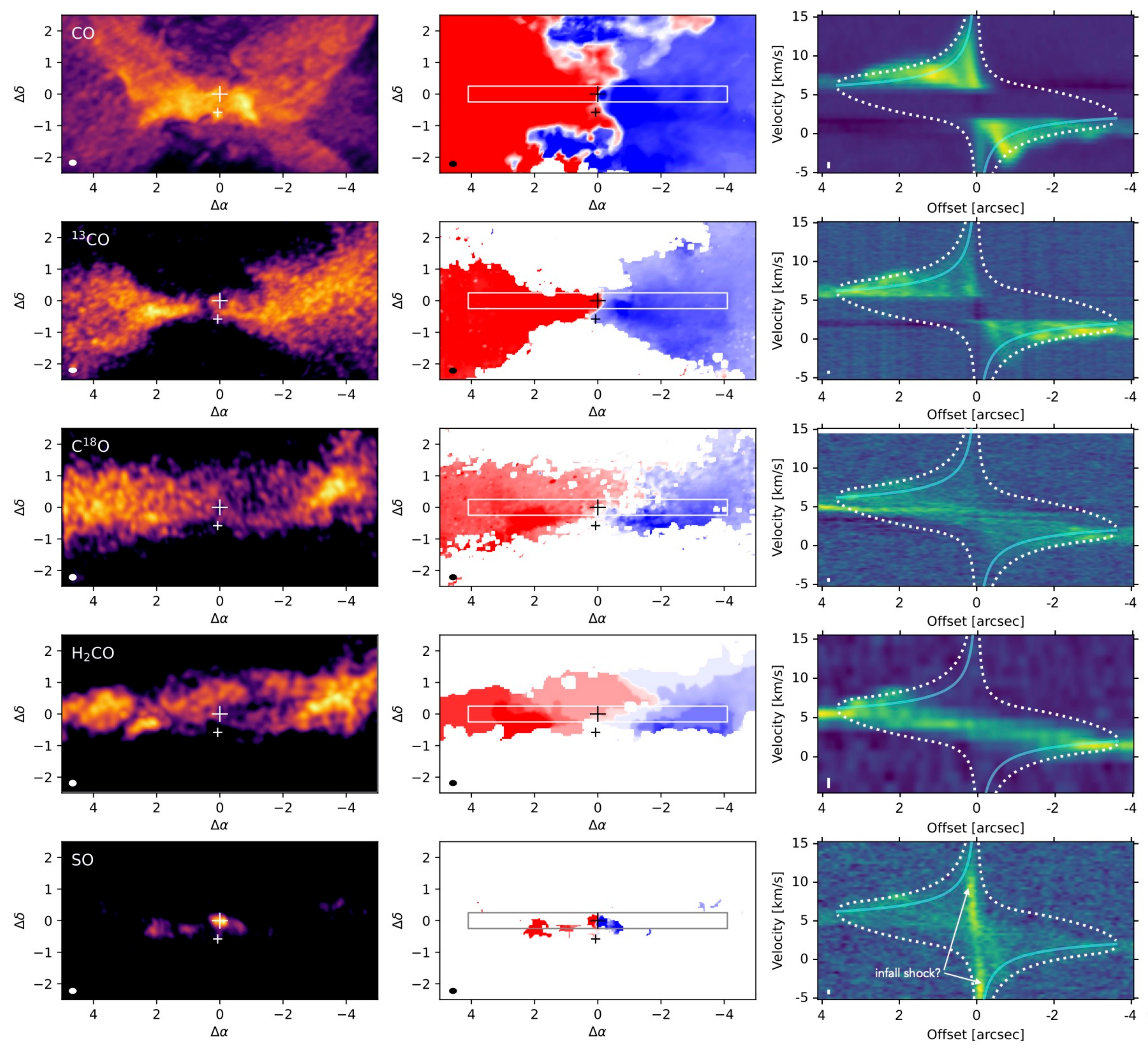}
\caption{Structure and kinematics of the IRS43 envelope as seen in the five brightest lines detected in our observations: the 2--1 transitions of CO, \13CO, and \C18O, H$_2$CO $3_{0,3}-2_{0,2}$, and SO $6_5-5_4$. The resolution of the data is indicated in the lower left corner of each panel.
The left column shows the peak emission map on an Asinh scale over the following ranges in mJy\,beam\e;
CO: 10--220; \13CO: 15--90; \C18O: 10--30; H$_2$CO: 5--15; SO: 10--30.
The large and small crosses indicate the positions of VLA1 and VLA2, respectively.
The middle column is the first moment for each line on the same velocity scale from 0 (red) to 6\,\kms\ (blue) and shows the East-West velocity gradient in the envelope. The rectangular box shows the cut along the RA axis with a width of 0\farcs5 in Dec used for the position-velocity diagrams displayed in the right column. All 5 tracers show emission at velocities greater than expected for pure Keplerian velocity for an edge-on geometry and a central mass of $2.5\,M_\odot$ (the maximum mass allowed by the astrometry, shown by the solid cyan line). The dotted white line brackets the range of projected velocities for a combination of rotation and free-fall collapse with the same central mass and matches the data much better.
The prominent SO emission centered on VLA1 has a high velocity gradient and may be due to an infall shock onto the disk.
}
\label{fig:pv}
\end{figure*}

The envelope shows a clear velocity gradient, redshifted to blueshifted from east to west centered on $\sim 4$\,\kms. A collage of peak emission, first moment, and position-velocity diagrams for the five brightest lines is shown in Figure~\ref{fig:pv}. The optically thick CO emission is the most spatially distributed and shows the temperature structure of the gas, highlighting the rims of the cavity as noted above.
The average brightness temperature over the envelope is 25\,K but the cavity rims are about a factor of two higher.
The \13CO\ emission is more optically thin and highlights the envelope, with the peak emission here showing a flared appearance.
The less abundant and optically thinner \C18O\ line is detected where the CO and \13CO\ lines are filtered out which provides an important view of the structure and kinematics of the envelope near the source velocity. The peak emission map is more uniform and less tightly pinched than the \13CO but also centered around VLA1, and the velocity gradient in the position-velocity map extends linearly through the center.
The H$_2$CO transition shown here was observed at relatively low velocity resolution, 1.34\,\kms, but is also less affected by spatial filtering than CO and \13CO\ near the central velocities and has a similar spatial and kinematic appearance as \C18O. Finally, the SO line was also weakly detected in the envelope but, unlike the other tracers, peaked strongly on VLA1.

The position-velocity diagrams shown in the third column of Figure~\ref{fig:pv} show that the velocity gradient across the envelope is not simply inherited shear or rotation of the molecular cloud or filament from which the stars formed, but increases around the stars due to their gravity.
The binary orbit has been accurately determined from the two decades of high resolution VLA and ALMA astrometry and the total mass of the system is constrained to be $M_\ast = 2.01\pm 0.47\,M_\odot$ \citep{Brinch2016}. However, the gas motions are significantly faster than Keplerian even for the maximum projected velocity case of the rotation axis being in the plane of the sky (solid cyan lines in Figure \ref{fig:pv}) and a central mass equal to the upper limit of $2.5\,M_\odot$. This therefore indicates that the envelope is not purely rotating.

Following \cite{Cesaroni2011} who considered the motions in an envelope around a young massive star, we include a free-fall component perpendicular to the rotation, which results in a radial velocity profile,

\begin{equation}
v(r) = \left(\frac{GM_\ast}{r}\right)^{1/2}\biggl(\frac{x}{r}+\sqrt2\frac{z}{r}\biggr), 
\end{equation}
where $x$ is the physical distance in the East-West direction in the plane of the sky, $z$ is the physical distance along the line-of-sight, such that the radius from the center of mass of the binary is $r = (x^2+z^2)^{1/2}$. The dotted white lines in Figure \ref{fig:pv} show the allowed range (minimum to maximum) of projected velocities that we would expect from this rotating and infalling model for the highest stellar mass consistent with the astrometry, $M_\ast=2.5\,M_\odot$. This matches the outer extent of the emission for each line in  Figure \ref{fig:pv} where the pure-Keplerian fit is too low. In addition, there is some flow toward the observer that lie within the expected bounds of the model for the less optically thin lines, \C18O, H$_2$CO, and SO, where we can see through to the back side.
Together, this demonstrates that the envelope is not only rotating but continuing to fall onto the system and the gravitational influence of the protostars is felt to beyond an outer radius $R>2500$\,au, limited by the map size and sensitivity of the data.

Only the SO line actually peaks on the source, and even then only on VLA1, not VLA2. The position-velocity diagram shows a high velocity gradient across VLA1, though 
it does not exceed the velocity range of the rotating-infalling envelope model.
Inspection of channel maps shows that the gradient follows the same East-West direction as the larger scale envelope. The weak, distributed SO emission in the envelope is similar to that seen in other embedded protostars \citep{Tychoniec2021}, but the strong enhancement on the source may signpost infall from the envelope onto the disk similar to observations by \citet{Sakai2014} and modeled by \citet{vanGelder2021}.
The velocity gradient of this feature is approximately perpendicular to the major axis of the dust disk so it is also possible that it may be a small outflow or jet.

\subsection{Signature of a recent outburst}\label{sec:bubble}
Understanding protostellar feedback and how newborn stars clear their surroundings is a key question in star formation and learning about the origin of the stellar mass function.
Neither the CO, SO or any other line have high velocity line wings indicative of an unbound outflow. Moreover, we do not detect SiO 5--4 in the passband, although it is detected in the powerful jets from several other eDisk sources.
It is therefore unclear what is the cause of the $\sim 10^3$\,au cavity blown out of the envelope. However, inspection of the CO channel maps reveal a possible clue to the recent accretion and outburst history of IRS43.

\begin{figure*}[ht]
\centering
\includegraphics[width=\textwidth]{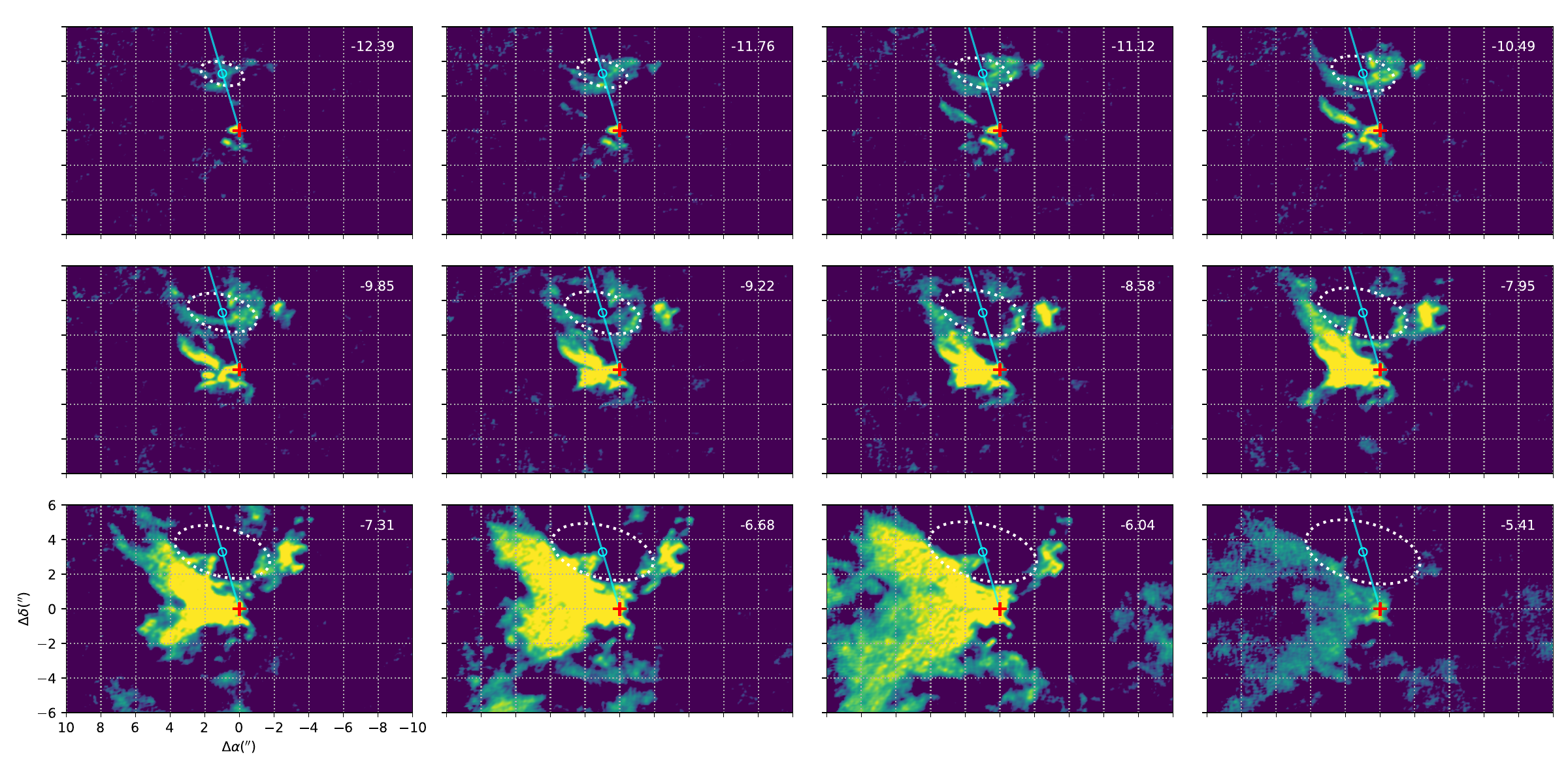}
\caption{Channel maps of CO for the red-shifted side of its emission. The velocity of each channel is indicated at the top right of each panel in units of \kms. The maps are centered on VLA1 (red cross) and the cyan line shows the proper motion of the source projected back in time. The white dashed ellipses, which have the same inclination and center (cyan open circle) and radii that scale linearly with velocity, outline a ring-like feature in each channel. We interpret this as an expanding bubble centered on the location of the binary 130 years ago.}
\label{fig:bubble}
\end{figure*}

Figure~\ref{fig:bubble} presents channel maps for the redshifted side of the envelope emission. We find
a ring like structure with a common spatial center but varying radius at different velocities (linearly proportional to the relative velocity difference) outlined by white dashed ellipses in each subplot. Remarkably, this ring center lies at the extrapolated position of the binary 130 years ago, as determined from its proper motion, $\mu_\alpha=-7.6$, $\mu_\delta=-25.3$\,mas year\e\
\citep{Brinch2016},
and illustrated by the solid blue line and open circle. We expect the molecular envelope is co-moving with the stars but, if something disrupted the surrounding gas, there could be features that decouple from this common motion. Consequently, we interpret this CO feature as an expanding molecular ring that is the remnant of a protostellar outburst event that occurred at the end of the 19$^{\rm th}$ century.

We can make a rough estimate of the energy of the bubble from its size, $\sim 500$\,au in radius, and expansion speed, $\sim 500\,{\rm au}/130\,{\rm yr} = 20$\,\kms.
If the density of the pre-burst gas was $n_{\rm H_2}=10^3$\,\cc, then the mass of gas in the bubble is about $1\,M_\oplus$ and its total kinetic energy is $\sim 2\times 10^{40}$\,erg.
This is about an order of magnitude greater than the total energy of the X-ray superflare observed by \citet{Grosso1997}, $L_{\rm X}\sim 10^{1.5}\,L_\odot$ over a couple of hours.
However, for a typical mechanical efficiency of a few percent for converting from stellar outburst scales to the ISM, the total energy requirement is perhaps similar to an EXor event but less than a FUor \citep{Hartmann_accretion_review}.

\section{Discussion}\label{sec:discussion}
The two disks around VLA1 and VLA2 are remarkably compact, just a few au in radius. They are the smallest in the eDisk sample \citep{Ohashi_edisk_overview} and may be the smallest yet measured with ALMA. It is well known that binary systems can severely truncate disks \citep{Artymowicz_1994} but the 74\,au separation of the two sources appears to be too large for this to be the explanation. For the $\sim 3$\,au disk around VLA2 to be dynamically affected, the semi-minor axis of the binary orbit would have to be no greater than 9\,au which implies a very high orbital eccentricity greater than 0.88. Interestingly, \citet{2019A&A...628A..95M} came to a similar conclusion for disks in multiple systems in Taurus. In this case, however, the positions of two sources have been accurately measured over 25\,yr (5\% of the orbital period) and is consistent with a circular orbit \citep{Brinch2016}.

ALMA surveys of Class II objects demonstrate that many dust disks are quite small, $\simlt 20$\,au in radius \citep{Manara_PP7}. A possible explanation is the radial drift of millimeter and larger sized grains with high Stokes numbers and a lack of pressure bumps to resist it \citep{2019A&A...626L...2F}. Drift timescales are very short so this could indeed be important for protostellar disks \citep{2020ApJ...905..162T}.

The high brightness temperature in VLA1 shows that the continuum emission is likely optically thick (Figure~\ref{fig:continuum_binary}) and its mass may well be significantly underestimated. VLA2 is about 10 times fainter so the optical depth correction may be small and it may well have low mass. However, this is hard to reconcile with its high accretion rate \citep{Greene2002}. Infall from the envelope and rapid transport through the disk may explain this. Alternatively, if most of the mass were extremely centrally concentrated, into a $\sim 1$\,au radial region, the emission would be both optically thick and beam diluted to a low brightness temperature in the $\sim 0\farcs05$ beam.

Infrared observations show that the high visual extinction to VLA2, $A_{\rm V}=40$\,mag, is much greater than the inferred value from the mean column density of the large scale circumbinary envelope, $N_{\rm H_2}\sim 10^{20}$\,cm\ee\ ($A_{\rm V}\simeq 0.2$\,mag), which is additional evidence for localized dust around the protostar. Moreover, strong CO ice absorption at $4.67\,\mu$m reveals the presence of dense, cold molecular gas enveloping both stars \citep{2011A&A...533A.112H}.


A long standing problem in star formation is the low average luminosity of protostars despite the need to gain of order a solar mass in a million years \citep{Kenyon_Hartmann_1995}. Punctuated bursts of accretion are a potential solution and a handful of protostars have been observed to dramatically brighten due to such an event \citep{Hartmann_accretion_review}. Although stars may indeed grow ``when we are not looking'', here we have found that associated outburst events may leave detectable signatures in the structure and kinematics of their surroundings. This was only possible due to the long-term astrometric monitoring of this binary system and consequent accurate measurements of its proper motion. Now, a decade into the ALMA era, there is potential to measure protostellar proper motions for other sources and search for fossil interactions with their surroundings.

It is unclear what caused the outburst 130 years ago. It does not appear to be a close passage by the other member of the binary since the orbital period is much longer, $\sim 450$\,yr, although it is worth considering that the system is part of a moderately clustered region where star formation is influenced by interactions and much more dynamic than the classical picture of isolated core collapse \citep{2012MNRAS.419.3115B}.
Rather, it is likely related to the highly active nature of one of the individual stars, as witnessed by the recent X-ray superflare. The analysis here was limited by the small disk sizes and lack of detectable lines in the ALMA data, but the system has strong CO ro-vibrational line emission and absorption in the M-band at 4\,$\mu$m \citep{2011A&A...533A.112H} which may provide an alternative means to study the disk properties.

The eDisk survey of 12 Class 0 and 7 Class I objects revealed a diverse set of disk properties \citep{Ohashi_edisk_overview}. Ranked by bolometric temperature, IRS43 is one of the more evolved sources but has the smallest disks in the sample. It is one of four binary systems and has the smallest projected separation though not by a significant margin. The circumstellar envelope surrounding the source is punctured by a large cavity but there are no signs of molecular outflows from either of the two protostars, in contrast to the rest of the sources in the survey. IRS43 mostly stands out due to its compact disks and high X-ray activity, properties that might possibly be related.

\section{Summary}\label{sec:summary}
These observations of Oph\,IRS43, a Class I protobinary system, are part of a homogeneous dataset of 19 embedded protostellar disks in the eDisk ALMA Large Program. Our main results are as follows:

\begin{itemize}
\item We detect disks around each member of the IRS43 binary in millimeter continuum emission. The northern source, VLA1, is about 10 times brighter than the southern source, VLA2, opposite to how they appear in the infrared. Both disks are extremely small, just a few au in radius, and masses are low but, especially for VLA1, likely underestimated due to high optical depth. A third disk is detected around the Class II source, GY 263, that lies within in the ALMA field-of-view and is found to be a transition disk with a $\sim 20$\,au radius cleared central cavity.

\item We map a flattened circumbinary envelope over $\sim 2000$\,au in the East-West direction in the dust continuum and multiple molecular species, \13CO, \C18O, H$_2$CO, and SO. The CO emission extends in the North-South direction and delineates the rim of a wide cavity. The envelope is moving faster than expected for Keplerian rotation and we deduce that the motions must include a component of infall onto the system. The shock as it hits the disk may explain the enhanced SO emission toward VLA1.

\item We discovered an expanding ring of CO emission in the channel maps with a center that sits at the projected position of the system 130 years ago. We interpret this as the signature of a fossil outburst with an energy estimated to be greater than the X-ray superflare observed in 1995 but lower than FU Ori type events.
\end{itemize}

The eDisk survey has revealed a wide range of disk properties in embedded protostellar systems. Here, we have presented the results from a first look at the data for one individual system. Future work will include looking for trends across the sample. However, it is already clear that disks form and evolve in quite heterogeneous ways. This has ramifications for studies of subsequent stages from the T Tauri Class II phase all the way to exoplanets. Understanding the nature of survey outliers such as the compact disks in IRS43 will be an important part of a complete picture of star and planet formation.

\vspace{5mm}
This paper makes use of the following ALMA data: ADS/JAO.ALMA\#2019.1.00261.L.
ALMA is a partnership of ESO (representing its member states), NSF (USA) and NINS (Japan), together with NRC (Canada), NSTC and ASIAA (Taiwan) and KASI (Republic of Korea), in cooperation with the Republic of Chile. The Joint ALMA Observatory is operated by ESO, AUI/NRAO and NAOJ. The National Radio Astronomy Observatory is a facility of the National Science Foundation operated under cooperative agreement by Associated Universities, Inc.
Funding support is acknowledged from
NSF GRFP grant No. 2236415 and NSF AST-2107841 (SN),
NSF AST-2107841 (JPW),
NASA RP 80NSSC22K1159 (JJT),
NSF AST-2108794 (LWL),
NSTC 109-2112-M-001-051, 110- 2112-M-001-031 (NO, CF),
NASA 80NSSC20K0533 and NSF AST-1910106 (ZYL),
Independent Research Fund Denmark grant No. 0135-00123B (JKJ, RS), 
PID2020-114461GB-I00 funded by MCIN/AEI/10.13039/501100011033 (IdG),
NSTC 110-2628-M-001-003-MY3 from the Taiwan National Science and Technology Council and AS-CDA-111-M03 from the Academia Sinica Career Development Award (H-WY).

\software{CASA (McMullin et al. 2007), astropy (The Astropy Collaboration 2013, 2018).}

\bibliographystyle{aasjournal}
\bibliography{refs}

\begin{thebibliography}{}
\expandafter\ifx\csname natexlab\endcsname\relax\def\natexlab#1{#1}\fi
\providecommand{\url}[1]{\href{#1}{#1}}
\providecommand{\dodoi}[1]{doi:~\href{http://doi.org/#1}{\nolinkurl{#1}}}
\providecommand{\doeprint}[1]{\href{http://ascl.net/#1}{\nolinkurl{http://ascl.net/#1}}}
\providecommand{\doarXiv}[1]{\href{https://arxiv.org/abs/#1}{\nolinkurl{https://arxiv.org/abs/#1}}}

\bibitem[{{Allen} {et~al.}(2002){Allen}, {Myers}, {Di Francesco}, {Mathieu},
  {Chen}, \& {Young}}]{Allen2002}
{Allen}, L.~E., {Myers}, P.~C., {Di Francesco}, J., {et~al.} 2002, \apj, 566,
  993, \dodoi{10.1086/338128}

\bibitem[{{Andrews} {et~al.}(2018){Andrews}, {Huang}, {P{\'e}rez}, {Isella},
  {Dullemond}, {Kurtovic}, {Guzm{\'a}n}, {Carpenter}, {Wilner}, {Zhang}, {Zhu},
  {Birnstiel}, {Bai}, {Benisty}, {Hughes}, {{\"O}berg}, \& {Ricci}}]{DSHARP}
{Andrews}, S.~M., {Huang}, J., {P{\'e}rez}, L.~M., {et~al.} 2018, \apjl, 869,
  L41, \dodoi{10.3847/2041-8213/aaf741}

\bibitem[{{Artymowicz} \& {Lubow}(1994)}]{Artymowicz_1994}
{Artymowicz}, P., \& {Lubow}, S.~H. 1994, \apj, 421, 651,
  \dodoi{10.1086/173679}

\bibitem[{{Bate}(2012)}]{2012MNRAS.419.3115B}
{Bate}, M.~R. 2012, \mnras, 419, 3115, \dodoi{10.1111/j.1365-2966.2011.19955.x}

\bibitem[{{Beckwith} {et~al.}(1990){Beckwith}, {Sargent}, {Chini}, \&
  {Guesten}}]{beckwithB}
{Beckwith}, S. V.~W., {Sargent}, A.~I., {Chini}, R.~S., \& {Guesten}, R. 1990,
  \aj, 99, 924, \dodoi{10.1086/115385}

\bibitem[{{Brinch} \& {J{\o}rgensen}(2013)}]{Brinch2013}
{Brinch}, C., \& {J{\o}rgensen}, J.~K. 2013, \aap, 559, A82,
  \dodoi{10.1051/0004-6361/201322463}

\bibitem[{{Brinch} {et~al.}(2016){Brinch}, {J{\o}rgensen}, {Hogerheijde},
  {Nelson}, \& {Gressel}}]{Brinch2016}
{Brinch}, C., {J{\o}rgensen}, J.~K., {Hogerheijde}, M.~R., {Nelson}, R.~P., \&
  {Gressel}, O. 2016, \apjl, 830, L16, \dodoi{10.3847/2041-8205/830/1/L16}

\bibitem[{{Cesaroni} {et~al.}(2011){Cesaroni}, {Beltr{\'a}n}, {Zhang},
  {Beuther}, \& {Fallscheer}}]{Cesaroni2011}
{Cesaroni}, R., {Beltr{\'a}n}, M.~T., {Zhang}, Q., {Beuther}, H., \&
  {Fallscheer}, C. 2011, \aap, 533, A73, \dodoi{10.1051/0004-6361/201117206}

\bibitem[{{Curiel} {et~al.}(2003){Curiel}, {Girart}, {Rodr{\'\i}guez}, \&
  {Cant{\'o}}}]{curiel}
{Curiel}, S., {Girart}, J.~M., {Rodr{\'\i}guez}, L.~F., \& {Cant{\'o}}, J.
  2003, \apjl, 582, L109, \dodoi{10.1086/367631}

\bibitem[{{Duch{\^e}ne} {et~al.}(2007){Duch{\^e}ne}, {Bontemps}, {Bouvier},
  {Andr{\'e}}, {Djupvik}, \& {Ghez}}]{2007A&A...476..229D}
{Duch{\^e}ne}, G., {Bontemps}, S., {Bouvier}, J., {et~al.} 2007, \aap, 476,
  229, \dodoi{10.1051/0004-6361:20077270}

\bibitem[{{Facchini} {et~al.}(2019){Facchini}, {van Dishoeck}, {Manara},
  {Tazzari}, {Maud}, {Cazzoletti}, {Rosotti}, {van der Marel}, {Pinilla}, \&
  {Clarke}}]{2019A&A...626L...2F}
{Facchini}, S., {van Dishoeck}, E.~F., {Manara}, C.~F., {et~al.} 2019, \aap,
  626, L2, \dodoi{10.1051/0004-6361/201935496}

\bibitem[{{Girart} {et~al.}(2000){Girart}, {Rodr{\'\i}guez}, \&
  {Curiel}}]{Girart2000}
{Girart}, J.~M., {Rodr{\'\i}guez}, L.~F., \& {Curiel}, S. 2000, \apjl, 544,
  L153, \dodoi{10.1086/317302}

\bibitem[{{Greene} \& {Lada}(2002)}]{Greene2002}
{Greene}, T.~P., \& {Lada}, C.~J. 2002, \aj, 124, 2185, \dodoi{10.1086/342861}

\bibitem[{{Grosso} {et~al.}(1997){Grosso}, {Montmerle}, {Feigelson},
  {Andr{\'e}}, {Casanova}, \& {Gregorio-Hetem}}]{Grosso1997}
{Grosso}, N., {Montmerle}, T., {Feigelson}, E.~D., {et~al.} 1997, \nat, 387,
  56, \dodoi{10.1038/387056a0}

\bibitem[{{Hartmann} {et~al.}(2016){Hartmann}, {Herczeg}, \&
  {Calvet}}]{Hartmann_accretion_review}
{Hartmann}, L., {Herczeg}, G., \& {Calvet}, N. 2016, \araa, 54, 135,
  \dodoi{10.1146/annurev-astro-081915-023347}

\bibitem[{{Herczeg} {et~al.}(2011){Herczeg}, {Brown}, {van Dishoeck}, \&
  {Pontoppidan}}]{2011A&A...533A.112H}
{Herczeg}, G.~J., {Brown}, J.~M., {van Dishoeck}, E.~F., \& {Pontoppidan},
  K.~M. 2011, \aap, 533, A112, \dodoi{10.1051/0004-6361/201016246}

\bibitem[{{J{\o}rgensen} {et~al.}(2009){J{\o}rgensen}, {van Dishoeck},
  {Visser}, {Bourke}, {Wilner}, {Lommen}, {Hogerheijde}, \& {Myers}}]{2009jes}
{J{\o}rgensen}, J.~K., {van Dishoeck}, E.~F., {Visser}, R., {et~al.} 2009,
  \aap, 507, 861, \dodoi{10.1051/0004-6361/200912325}

\bibitem[{{Kenyon} \& {Hartmann}(1995)}]{Kenyon_Hartmann_1995}
{Kenyon}, S.~J., \& {Hartmann}, L. 1995, \apjs, 101, 117,
  \dodoi{10.1086/192235}

\bibitem[{{Manara} {et~al.}(2022){Manara}, {Ansdell}, {Rosotti}, {Hughes},
  {Armitage}, {Lodato}, \& {Williams}}]{Manara_PP7}
{Manara}, C.~F., {Ansdell}, M., {Rosotti}, G.~P., {et~al.} 2022, arXiv
  e-prints, arXiv:2203.09930, \dodoi{10.48550/arXiv.2203.09930}

\bibitem[{{Manara} {et~al.}(2019){Manara}, {Tazzari}, {Long}, {Herczeg},
  {Lodato}, {Rota}, {Cazzoletti}, {van der Plas}, {Pinilla}, {Dipierro},
  {Edwards}, {Harsono}, {Johnstone}, {Liu}, {Menard}, {Nisini}, {Ragusa},
  {Boehler}, \& {Cabrit}}]{2019A&A...628A..95M}
{Manara}, C.~F., {Tazzari}, M., {Long}, F., {et~al.} 2019, \aap, 628, A95,
  \dodoi{10.1051/0004-6361/201935964}

\bibitem[{{McMullin} {et~al.}(2007){McMullin}, {Waters}, {Schiebel}, {Young},
  \& {Golap}}]{McMullin2007}
{McMullin}, J.~P., {Waters}, B., {Schiebel}, D., {Young}, W., \& {Golap}, K.
  2007, in Astronomical Society of the Pacific Conference Series, Vol. 376,
  Astronomical Data Analysis Software and Systems XVI, ed. R.~A. {Shaw},
  F.~{Hill}, \& D.~J. {Bell}, 127

\bibitem[{{Montmerle} {et~al.}(2000){Montmerle}, {Grosso}, {Tsuboi}, \&
  {Koyama}}]{Montmerle2000}
{Montmerle}, T., {Grosso}, N., {Tsuboi}, Y., \& {Koyama}, K. 2000, \apj, 532,
  1097, \dodoi{10.1086/308611}

\bibitem[{{Offner} {et~al.}(2010){Offner}, {Kratter}, {Matzner}, {Krumholz}, \&
  {Klein}}]{Offner2010}
{Offner}, S. S.~R., {Kratter}, K.~M., {Matzner}, C.~D., {Krumholz}, M.~R., \&
  {Klein}, R.~I. 2010, \apj, 725, 1485, \dodoi{10.1088/0004-637X/725/2/1485}

\bibitem[{{Ohashi} {et~al.}(2023){Ohashi}, {Tobin}, {J{\o}rgensen}, {Takakuwa},
  {Sheehan}, {Aikawa}, {Li}, {Looney}, {Williams}, {Aso}, {Sharma}, {Sai},
  {Yamato}, {Lee}, {Tomida}, {Yen}, {Encalada}, {Flores}, {Gavino}, {Kido},
  {Han}, {Lin}, {Narayanan}, {Phuong}, {Santamar{\'\i}a-Miranda}, {Thieme},
  {van't Hoff}, {de Gregorio-Monsalvo}, {Koch}, {Kwon}, {Lai}, {Lee},
  {Plunkett}, {Saigo}, {Hirano}, {Lam}, \& {Mori}}]{Ohashi_edisk_overview}
{Ohashi}, N., {Tobin}, J.~J., {J{\o}rgensen}, J.~K., {et~al.} 2023, \apj, 951,
  8, \dodoi{10.3847/1538-4357/acd384}

\bibitem[{{Ortiz-Le{\'o}n} {et~al.}(2017){Ortiz-Le{\'o}n}, {Loinard},
  {Kounkel}, {Dzib}, {Mioduszewski}, {Rodr{\'\i}guez}, {Torres},
  {Gonz{\'a}lez-L{\'o}pezlira}, {Pech}, {Rivera}, {Hartmann}, {Boden}, {Evans},
  {Brice{\~n}o}, {Tobin}, {Galli}, \& {Gudehus}}]{OrtizLeon2017}
{Ortiz-Le{\'o}n}, G.~N., {Loinard}, L., {Kounkel}, M.~A., {et~al.} 2017, \apj,
  834, 141, \dodoi{10.3847/1538-4357/834/2/141}

\bibitem[{{Raghavan} {et~al.}(2010){Raghavan}, {McAlister}, {Henry}, {Latham},
  {Marcy}, {Mason}, {Gies}, {White}, \& {ten
  Brummelaar}}]{Raghavan_binary_survey}
{Raghavan}, D., {McAlister}, H.~A., {Henry}, T.~J., {et~al.} 2010, \apjs, 190,
  1, \dodoi{10.1088/0067-0049/190/1/1}

\bibitem[{{Ripple} {et~al.}(2013){Ripple}, {Heyer}, {Gutermuth}, {Snell}, \&
  {Brunt}}]{2013MNRAS.431.1296R}
{Ripple}, F., {Heyer}, M.~H., {Gutermuth}, R., {Snell}, R.~L., \& {Brunt},
  C.~M. 2013, \mnras, 431, 1296, \dodoi{10.1093/mnras/stt247}

\bibitem[{{Sakai} {et~al.}(2014){Sakai}, {Sakai}, {Hirota}, {Watanabe},
  {Ceccarelli}, {Kahane}, {Bottinelli}, {Caux}, {Demyk}, {Vastel}, {Coutens},
  {Taquet}, {Ohashi}, {Takakuwa}, {Yen}, {Aikawa}, \& {Yamamoto}}]{Sakai2014}
{Sakai}, N., {Sakai}, T., {Hirota}, T., {et~al.} 2014, \nat, 507, 78,
  \dodoi{10.1038/nature13000}

\bibitem[{{Thebault} \& {Haghighipour}(2015)}]{2015pes..book..309T}
{Thebault}, P., \& {Haghighipour}, N. 2015, in Planetary Exploration and
  Science: Recent Results and Advances, 309--340,
  \dodoi{10.1007/978-3-662-45052-9_13}

\bibitem[{{Tobin} {et~al.}(2015){Tobin}, {Looney}, {Wilner}, {Kwon},
  {Chandler}, {Bourke}, {Loinard}, {Chiang}, {Schnee}, \& {Chen}}]{Tobin2015}
{Tobin}, J.~J., {Looney}, L.~W., {Wilner}, D.~J., {et~al.} 2015, \apj, 805,
  125, \dodoi{10.1088/0004-637X/805/2/125}

\bibitem[{{Tobin} {et~al.}(2020){Tobin}, {Sheehan}, {Reynolds}, {Megeath},
  {Osorio}, {Anglada}, {D{\'\i}az-Rodr{\'\i}guez}, {Furlan}, {Kratter},
  {Offner}, {Looney}, {Kama}, {Li}, {van't Hoff}, {Sadavoy}, \&
  {Karnath}}]{2020ApJ...905..162T}
{Tobin}, J.~J., {Sheehan}, P.~D., {Reynolds}, N., {et~al.} 2020, \apj, 905,
  162, \dodoi{10.3847/1538-4357/abc5bf}

\bibitem[{{Tychoniec} {et~al.}(2021){Tychoniec}, {van Dishoeck}, {van't Hoff},
  {van Gelder}, {Tabone}, {Chen}, {Harsono}, {Hull}, {Hogerheijde}, {Murillo},
  \& {Tobin}}]{Tychoniec2021}
{Tychoniec}, {\L}., {van Dishoeck}, E.~F., {van't Hoff}, M. L.~R., {et~al.}
  2021, \aap, 655, A65, \dodoi{10.1051/0004-6361/202140692}

\bibitem[{{van der Marel} {et~al.}(2022){van der Marel}, {Williams}, {Picogna},
  {van Terwisga}, {Facchini}, {Manara}, {Zormpas}, {Ansdell}, \&
  {.}}]{vanderMarel2022}
{van der Marel}, N., {Williams}, J.~P., {Picogna}, G., {et~al.} 2022, arXiv
  e-prints, arXiv:2204.08225.
\newblock \doarXiv{2204.08225}

\bibitem[{{van der Tak} {et~al.}(2007){van der Tak}, {Black}, {Sch{\"o}ier},
  {Jansen}, \& {van Dishoeck}}]{vanderTak2007}
{van der Tak}, F.~F.~S., {Black}, J.~H., {Sch{\"o}ier}, F.~L., {Jansen}, D.~J.,
  \& {van Dishoeck}, E.~F. 2007, \aap, 468, 627,
  \dodoi{10.1051/0004-6361:20066820}

\bibitem[{{van Gelder} {et~al.}(2021){van Gelder}, {Tabone}, {van Dishoeck}, \&
  {Godard}}]{vanGelder2021}
{van Gelder}, M.~L., {Tabone}, B., {van Dishoeck}, E.~F., \& {Godard}, B. 2021,
  \aap, 653, A159, \dodoi{10.1051/0004-6361/202141591}

\bibitem[{{Williams} \& {Best}(2014)}]{2014ApJ...788...59W}
{Williams}, J.~P., \& {Best}, W. M.~J. 2014, \apj, 788, 59,
  \dodoi{10.1088/0004-637X/788/1/59}

\end{thebibliography}

\end{document}